# Modeling the Worldwide Spread of Pandemic Influenza: Baseline Case and Containment Interventions

Vittoria Colizza[1,2], Alain Barrat[2,3], Marc Barthelemy[1,4], Alain-Jacques Valleron[5,6], Alessandro Vespignani[1,2*]

1 School of Informatics and Center for Biocomplexity, Indiana University, Bloomington, Indiana, United States of America, 2 Complex Networks Lagrange Laboratory, Institute for Scientific Interchange Foundation, Turin, Italy, 3 Unité Mixte de Recherche 8627, Centre National de la Recherche Scientifique, Université Paris-Sud, Orsay, France, 4 Commissariat à l'Energie Atomique, Direction Ile-de-France, Centre d'Etudes de Bruyères le Châtel, Bruyères le Châtel, France, 5 Unité de Recherche en Epidémiologie, Systèmes d'Information et Modélisation, Institut National de la Santé et de la Recherche Médicale, Unité Mixte de Recherche 707, Université Pierre et Marie Curie, Paris, France, 6 Hôpital Saint-Antoine, Assistance Publique-Hôpitaux de Paris, Paris, France

**Funding:** AB and AV are partially funded by the European Commission contract 001907 Dynamically Evolving Large-scale Information Systems (DELIS). AV is partially funded by the National Science Foundation award IIS-0513650. The funders had no role in study design, data collection and analysis, decision to publish, or preparation of the manuscript.

**Competing Interests:** The authors have declared that no competing interests exist.

**Academic Editor:** Alison P. Galvani, Yale University, United States of America

**Citation:** Colizza V, Barrat A, Barthelemy M, Valleron AJ, Vespignani A (2007) Modeling the worldwide spread of pandemic influenza: Baseline case and containment interventions. PLoS Med 4(1): e13. doi:10.1371/journal.pmed.0040013

**Received:** April 27, 2006
**Accepted:** November 27, 2006
**Published:** January 23, 2007

## A B S T R A C T

### Background

The highly pathogenic H5N1 avian influenza virus, which is now widespread in Southeast Asia and which diffused recently in some areas of the Balkans region and Western Europe, has raised a public alert toward the potential occurrence of a new severe influenza pandemic. Here we study the worldwide spread of a pandemic and its possible containment at a global level taking into account all available information on air travel.

### Methods and Findings

We studied a metapopulation stochastic epidemic model on a global scale that considers airline travel flow data among urban areas. We provided a temporal and spatial evolution of the pandemic with a sensitivity analysis of different levels of infectiousness of the virus and initial outbreak conditions (both geographical and seasonal). For each spreading scenario we provided the timeline and the geographical impact of the pandemic in 3,100 urban areas, located in 220 different countries. We compared the baseline cases with different containment strategies, including travel restrictions and the therapeutic use of antiviral (AV) drugs. We investigated the effect of the use of AV drugs in the event that therapeutic protocols can be carried out with maximal coverage for the populations in all countries. In view of the wide diversity of AV stockpiles in different regions of the world, we also studied scenarios in which only a limited number of countries are prepared (i.e., have considerable AV supplies). In particular, we compared different plans in which, on the one hand, only prepared and wealthy countries benefit from large AV resources, with, on the other hand, cooperative containment scenarios in which countries with large AV stockpiles make a small portion of their supplies available worldwide.

### Conclusions

We show that the inclusion of air transportation is crucial in the assessment of the occurrence probability of global outbreaks. The large-scale therapeutic usage of AV drugs in all hit countries would be able to mitigate a pandemic effect with a reproductive rate as high as 1.9 during the first year; with AV supply use sufficient to treat approximately 2% to 6% of the population, in conjunction with efficient case detection and timely drug distribution. For highly contagious viruses (i.e., a reproductive rate as high as 2.3), even the unrealistic use of supplies corresponding to the treatment of approximately 20% of the population leaves 30%–50% of the population infected. In the case of limited AV supplies and pandemics with a reproductive rate as high as 1.9, we demonstrate that the more cooperative the strategy, the more effective are the containment results in all regions of the world, including those countries that made part of their resources available for global use.

The Editors' Summary of this article follows the references.

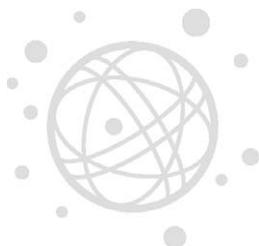

## Introduction

The avian influenza pathogen H5N1 has proven its ability to pass directly from birds to humans, causing a severe disease with high mortality [1]. This has raised a public alert toward the potential occurrence of a new severe influenza pandemic caused by a novel human virus strain originating from the avian H5N1 influenza virus, through adaptive mutations or recombinations with human influenza viruses. In this eventuality, the timescale necessary for the worldwide production and deployment of adequate vaccine supplies could exceed six to eight months [2]. In addition, in the case of diseases in which the infected individuals are infectious before the appearance of the first clinical signs [3], non-medical interventions such as travel-related measures might not be as efficient [4] as they proved to be during the spread of SARS [5]. In this context, antiviral (AV) drugs represent one of the crucial resources to reduce morbidity and mortality of a pandemic in the absence of a vaccine [6–9]. For this reason, several recent studies provided evidence that the use of AV drugs is an effective strategy for the local containment of the emerging pandemic strain at the source [7,10].

The real issue in assessing the impact of a pandemic is, however, in its global character, and in the effects resulting from the complex interplay of epidemic spread occurring in different countries. The study of the global scale of a pandemic must examine three major factors in modeling studies: (i) In learning from the lessons of the 1918 pandemic and the ongoing debate on its origins, scenario evaluations must investigate the possibility of a pandemic starting anywhere in the world and not only in underdeveloped countries. (ii) Determining the likelihood of the global spread of a pandemic must include consideration of the effects of human travel. Even if local intervention were to mitigate and eradicate an epidemic locally, it is possible that during the course of containment, which generally lasts for weeks, air traffic might trigger a global event affecting multiple countries. The air traffic would also alter the evolution of the local epidemics with the entry of new, infectious individuals from elsewhere, a nonlocal effect not specifically considered in previous studies [7,10], or implemented as assumed boundary conditions [11,12]. (iii) AV stockpiling is mainly a function of a country's wealth and level of preparedness to deal with a future pandemic; consequently, stockpiles are very low or absent in a majority of countries. It is, therefore, relevant to study effective strategies that optimize the use of the available supplies of AV drugs in the case of a worldwide spread.

In this article, we examine stochastic computational modeling of the temporal and worldwide geographical spread of an emerging pandemic [13,14]. The epidemic model includes census data for all 3,100 urban areas studied and airline traffic flow between those regions, all of which accounts for 99% of airline traffic worldwide. Such detailed information allows for the study of a pandemic originating anywhere in the world. We first investigate the pandemic evolution on a global scale in the absence of containment strategies, studying the effect of different initial conditions and virus infectiousness, quantified by the basic reproductive number $R_0$. This preliminary analysis allows us to outline a set of baseline cases to contrast with different intervention scenarios. We first analyse the effect of travel restrictions on the overall evolution of the pandemic, and then consider the best-case scenario in which each hit country can rely on massive AV stockpiles. We also investigate the effect of traveling patterns on the occurrence of a global outbreak affecting a large number of countries worldwide even if an effective mitigation of the epidemic is achieved in the hit populations. Finally, with the aim of designing realistic scenarios concerning AV stockpiles, we take into consideration that the amount of available supplies is finite and concentrated in a limited number of countries. We therefore contrast uncooperative intervention strategies, in which prepared countries use their stockpiles only within their own borders, against increasingly cooperative strategies in which progressively larger fractions of the stockpiles of prepared countries are shared with unprepared countries worldwide.

## Methods

### Socioeconomic Data

The International Air Transport Association (IATA) (http://www.iata.org) database contains the worldwide list of airport pairs connected by direct flights and the number of available seats on any given connection. The resulting worldwide air-transportation network is therefore a weighted graph comprising $V = 3,100$ vertices denoting airports in 220 different countries (see Figure 1), and $E = 17,182$ weighted edges whose weight $w_{jl}$ represents the passenger flow between the airports $j$ and $l$. This dataset accounts for 99% of worldwide traffic and has been complemented by the population $N_j$ of each large metropolitan area served by the corresponding airport as obtained from different sources. The network that was obtained is highly heterogeneous both in its connectivity pattern and traffic capacities [13,14].

### The Model

As a basic modeling strategy we used a metapopulation approach [15–17] in which individuals are allowed to travel from one city to another by means of the airline transportation network, and to change compartments because of the infection dynamics in each city (see Figure 1), similarly to the models elsewhere [18–23] and the stochastic generalizations in other studies [13,14,24]. Specifically, two studies [19,22] address the computational analysis of influenza pandemics. These approaches were limited, however, by a small dataset of 50 cities available at that time and a simplified compartmental model. Moreover, the methods [19,22] do not take into account stochastic effects and AV distribution, or other containment measures. In this study, we are in a position to take advantage of the recent increase in computer power and scale up the modeling approach to utilize the IATA dataset to its fullest extent, as well as taking the stochastic nature of the dynamics into account. We used the standard compartmentalization in which individuals can exist only in one of the discrete states such as susceptible *(S)*, latent *(L)*, infected *(I)*, permanently recovered *(R)*, etc. In the case of AV administration we used up to seven different compartments (see below). In each city $j$ the population is $N_j$, and $X_j^{[m]}(t)$ is the number of individuals in the state *(m)* at time $t$. By definition it follows that $N_j = \sum_m X_j^{[m]}(t)$. The dynamics of individuals based on travels between cities is described by

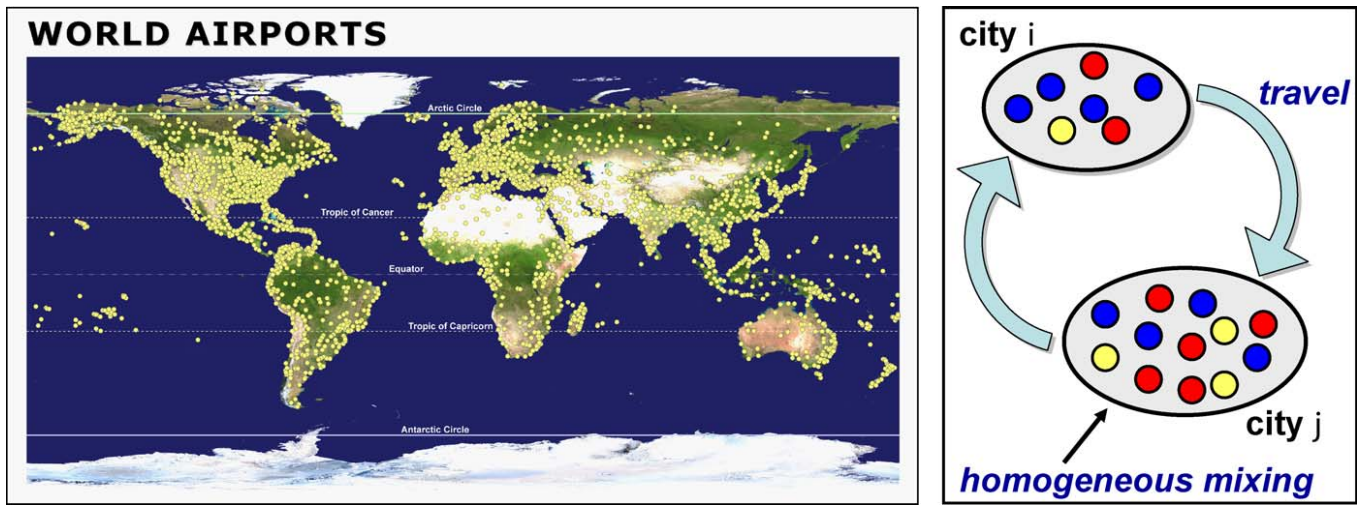

**Figure 1.** Representation of the Metapopulation Model Adopted

The model includes 3,100 airports in 220 countries worldwide. The map portrays the locations of urban airports worldwide; geographical data are obtained from open sources on the Internet and mapped with ArcGIS software (http://www.esri.com/software/arcgis). A schematic illustration represents the patch or metapopulation model adopted, in which the total population is divided into subpopulations each corresponding to the urban area surrounding each airport. Filled circles inside each subpopulation represent individuals, and the colors correspond to a specific stage of the disease. Homogeneous mixing for the infection dynamics is assumed inside each urban area, and different urban areas; subpopulations are coupled by means of air travel, according to the International Air Transport Association traffic fluxes.

doi:10.1371/journal.pmed.0040013.g001

the stochastic transport operator $\Omega_j(\{X^{[m]}\})$ representing the net balance of individuals in a given state $X^{[m]}$ that entered and left each city $j$. This operator is a function of the traffic flows with the neighboring cities $w_{jl}$ per unit time and of the city population $N_j$. In particular, the number of passengers in each category traveling from a city $j$ to a city $l$ is an integer random variable, in that each of the potential travelers has a probability $p_{jl} = w_{jl}\Delta t/N_j$ to go from $j$ to $l$ in the time interval $\Delta t$. In each city $j$ the numbers of passengers traveling on each connection $j \rightarrow l$ at time $t$ define a set of stochastic variables that follow a multinomial distribution [13,14]. The calculation can be extended to include transit traffic (e.g., up to one connection flight).

The infection evolution inside each urban area is described by compartmental schemes [6,7], in which the dynamics of the individuals among the different compartments depend on the specific etiology of the disease and the containment interventions considered. We used the two compartmental schemes reported in Figure 2. In the baseline scenario with no intervention, a susceptible individual (S) in contact with a symptomatic individual ($I^t$, $I^{nt}$) or asymptomatic ($I^a$) infectious individual becomes infected with rate $\beta$ or $r_\beta\beta$, respectively, and enters the latent class (L). When the latency period ends, the individual becomes infectious, i.e. able to transmit the infection, developing symptoms ($I^t$, $I^{nt}$) with probability $1 - p_a$, while becoming asymptomatic ($I^a$) with probability $p_a$. Among the symptomatic individuals, we distinguished between those who are allowed to travel ($I^t$)—with probability $p_t$—and those who are not allowed ($I^{nt}$)—with probability $1 - p_t$—depending on the severity of the disease. After the infectious period, all infectious individuals enter the recovered class (R). $\varepsilon^{-1}$ and $\mu^{-1}$ represent the mean latency period and the average duration of infection, respectively. We assumed that the latency period coincides with the incubation period with mean length of $\varepsilon^{-1} = 1.9$ d, followed by the infectious phase of average duration $\mu^{-1} = 3$ d [6–8]. Both the incubation and infectious periods are distributed exponentially around these averages. We assumed the probability of being asymptomatic given that infection has occurred to be $p_a = 33\%$ [6,7]. The relative infectiousness of asymptomatic individuals is $r_\beta = 50\%$. If the individual develops symptoms during the infectious period, he/she is then restricted from traveling ($I^{nt}$) with probability $1 - p_t = 50\%$ [6,7].

Whenever AV stockpiles are available, a certain fraction per day $p_{AV}$ of symptomatic infected individuals will enter the AV treatment, the efficacy of which is modeled through a reduction of both the infectiousness and the period during which the individual is contagious [6,7]. Moreover, individuals under AV treatment are not allowed to travel. Note that asymptomatic individuals do not look for health-care assistance and therefore cannot receive any treatment. The rate $p_{AV}$ takes into account the probability per unit time of detecting a symptomatic infected individual and the efficiency of AV distribution in the infected area. We assumed that ill individuals under treatment might naturally recover, thus modeling a possible excessive use of AV courses. The infectiousness of an ill individual under treatment is reduced by a factor $AVE_I = 0.62$ which represents the efficacy of the AV drugs [7]; in this case the transmission parameter is thus given by $\beta(1 - AVE_I)$. Moreover, the average infectious period for a treated ill individual is reduced by one day. The combination of $p_{AV}$, $AVE_I$, and the reduction of the average infectious period corresponds to a reduction of the overall infectiousness of treated individuals ranging from $50\%$ to $30\%$. In the Text S1 we report the sensitivity analysis for different overall reductions by changing the reduction of the infectious period. When AV stockpiles terminate in a given country, infected individuals are no longer treated (which is equivalent to set locally $p_{AV} = 0$).

Given the above compartments, in each city the epidemic is described by a set of stochastic differential equations (one for each compartment) obtained in the homogeneous assump-

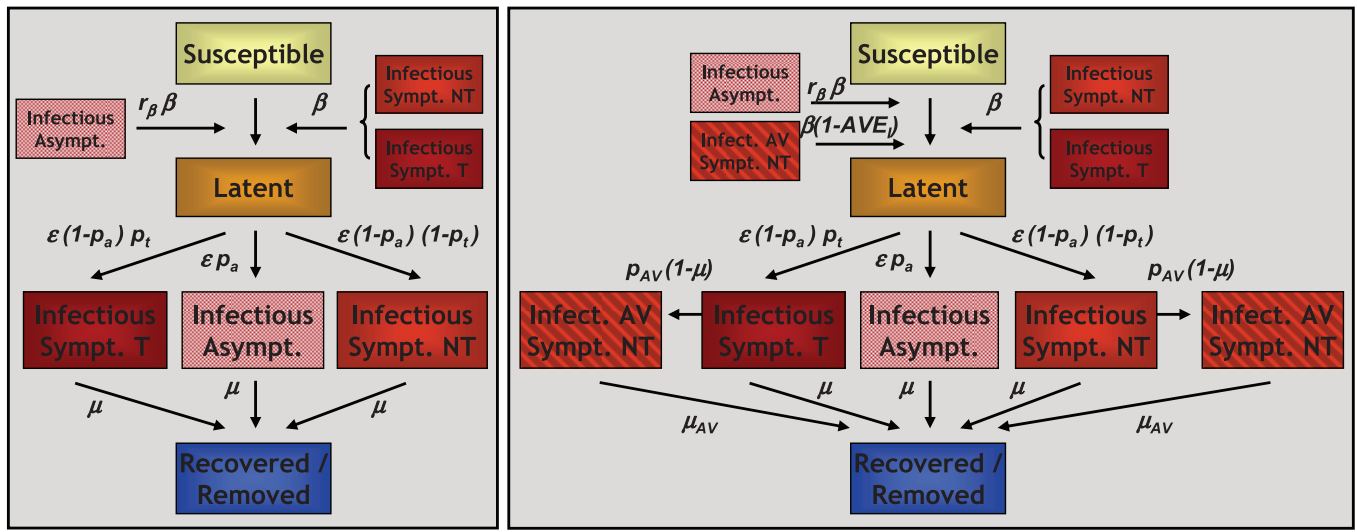

**Figure 2.** Flow Diagrams of the Transmission Models

The compartmentalization schemes adopted in the baseline scenario (left) and in the presence of intervention with the use of therapeutic AV treatment (right) are shown. "Sympt. T" and "Sympt. NT" correspond to infected symptomatic individuals allowed to travel and not allowed to travel, respectively. "Infect. AV" indicates infectious individuals under AV treatment, who are therefore not allowed to travel (NT).

doi:10.1371/journal.pmed.0040013.g002

tion with the addition of stochastic noise following the Langevin formulation and coupled to the population compartments in other cities by the stochastic transport operator. Here we report as an example the differential equation governing the evolution of the latent individuals in city $j$ in the baseline scenario

$$L_j(t + \Delta t) - L_j(t)$$

$$= \beta \frac{(I_j^t + I_j^{nt} + r_\beta I_j^a)S_j}{N_j}\Delta t - \varepsilon L_j \Delta t - \sqrt{\beta \frac{I_j^t S_j}{N_j}\Delta t}\, \eta_{\beta_t,j}(t) +$$

$$- \sqrt{\beta \frac{I_j^{nt} S_j}{N_j}\Delta t}\, \eta_{\beta_{nt},j}(t) - \sqrt{\beta \frac{r_\beta I_j^a S_j}{N_j}\Delta t}\, \eta_{\beta_a,j}(t) + \sqrt{\varepsilon L_j \Delta t}\, \eta_{\varepsilon,j}(t)$$

$$+ \Omega_j(\{L\}),$$

(1)

where $\eta_{\beta_t,j}$, $\eta_{\beta_{nt},j}$, $\eta_{\beta_a,j}$ and $\eta_{\varepsilon,j}$ are statistically independent Gaussian random variables with zero mean and unit variance. The term $\Omega_j$ represents the stochastic transport operator, reported in the Text S1 along with the full set of equations. The model therefore consists of a number of coupled evolution equations equal to the number of cities times the number of compartments considered. The epidemic evolution is obtained by solving numerically this set of coupled equations, using as initial condition a single symptomatic infected individual in a given urban area. The time step used in the solution of the equation corresponds to one day. The equations also consider the discrete nature of the individuals by using a specific calculation scheme reported in the Text S1. A fully discrete version of the model based on discrete multinomial extraction for all compartmental transitions recovers the same results. It is worth stressing that the only parameters in the models are the disease transition rates describing the etiology of the disease. All the couplings based on transportation terms, urban area size, and the network connectivity pattern are fixed elements of the problem obtained from real data.

## Reproductive Number $R_0$

The key parameter for the description of the spread rate of a disease is represented by the basic reproductive number $R_0$, defined as the average number of infections a typical infectious individual can generate in a fully susceptible population [25]. By computing the eigenvalues of the Jacobian at the disease-free equilibrium [26], we obtained the following expression for the basic reproductive number of the adopted compartmentalization: $R_0 = \frac{\beta}{\mu}(r_\beta p_a + 1 - p_a)$. The reproductive number used in the following discussion and reported in the figures refers to the value obtained by the previous expression and discounts all effects due to seasonality and AV interventions. An estimate of the reproductive number in the case of AV interventions is reported in the Text S1.

## Seasonality

To take into account the high influenza incidence during the winter and low incidence during the summer, we included seasonality in the infection transmission of the model. We followed the approach described in references [19,22,23] by using a transmission parameter $\beta$, which varies in time and depends on the geographic zone. To each city, we assigned a scaling factor for the transmission parameter, which is based on both the time of the year and the city's climate zone. We report the complete table of scaling factors and the geographic zone classification adopted in the Text S1.

## Results

### Baseline Cases

We first inspected the baseline results obtained under the hypothesis of no specific intervention to contain the outbreak. We considered four scenarios for the basic reproduction rate of the pandemic, namely $R_0 = 1.1$, $R_0 = 1.5$, $R_0 = 1.9$, and $R_0 = 2.3$. These values account for a broad spectrum of infectiveness of the virus ranging from a very mild situation (due to a potential loss of transmissibility with respect to the

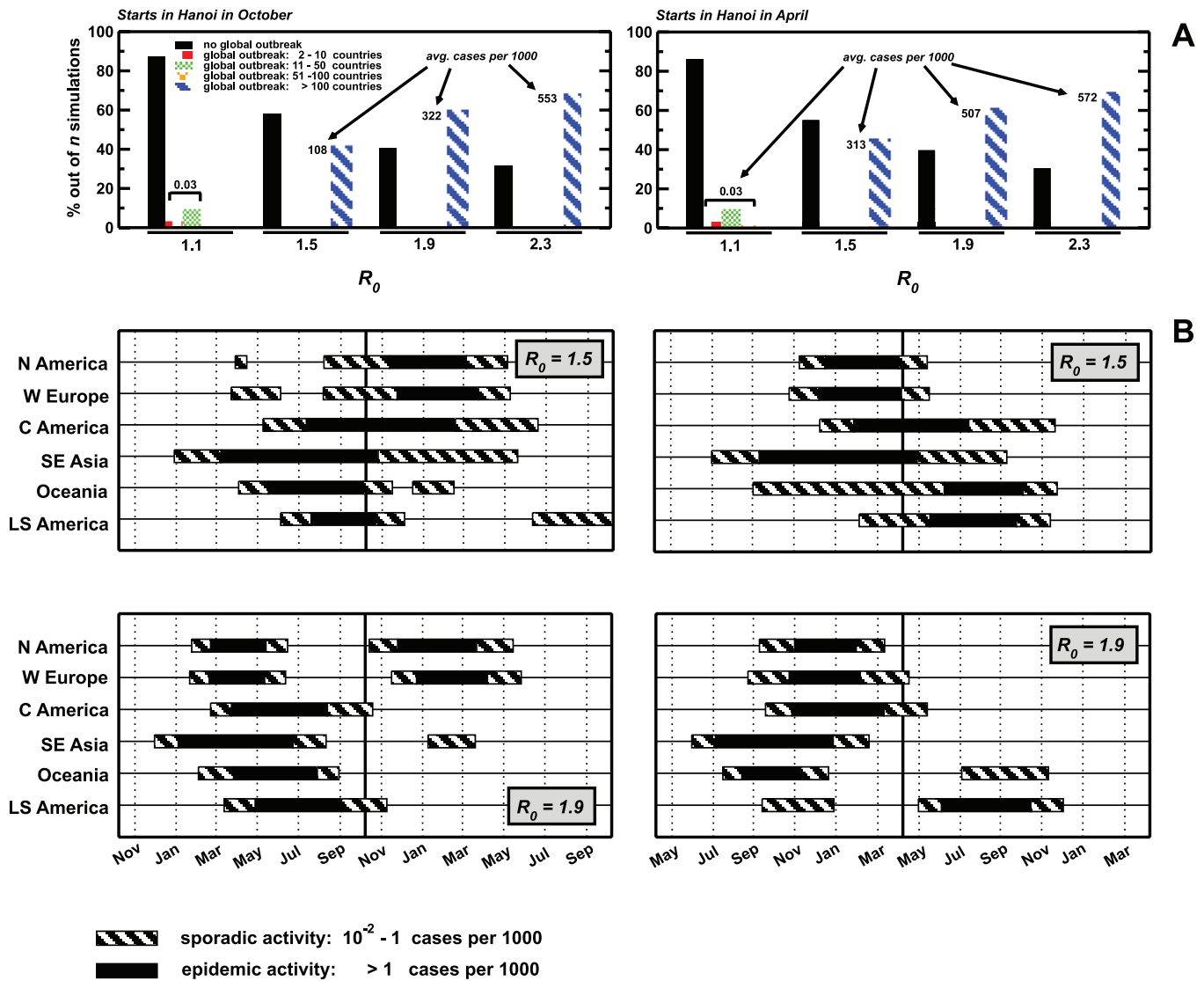

**Figure 3.** Baseline Scenario: Probability of Global Outbreaks and Expected Evolution for a Pandemic Seeded in Hanoi

(A) Probability of observing a global outbreak after one year from the beginning of the epidemic, for the values $R_0 = 1.1$, 1.5, 1.9, and 2.3. A global outbreak occurs when more than one country reports infected cases, while the absence of a global outbreak is given by an epidemic affecting the initially seeded country only. Two different dates for the start of the pandemic are shown: October (left) and April (right). The average attack rates in the case that a global outbreak occurs are also reported, by showing the average number of cases per 1,000 at the end of the first year. Results are obtained from $n = 500$ different realizations of the noise.

(B) Timing of the influenza pandemic in the six regions under study, for $R_0 = 1.5$ and $R_0 = 1.9$, for two different initial dates: October (left column) and April (right column). Sporadic activity occurs when the average prevalence reports ($10^{-2} - 1$) cases per 1,000, while epidemic activity is defined by more than one case per 1,000. The average profile is obtained by considering all runs for each region that experienced an outbreak in the region itself during the time window investigated. The thick vertical line corresponds to one year after the start of the pandemic.
doi:10.1371/journal.pmed.0040013.g003

avian influenza virus [4]) to values comparable to the 1918 pandemic [27]. While estimates of the reproductive rate for pandemic influenza vary considerably in the literature [3,8,19,28,29], recent studies tend to revise downward large values of $R_0$ previously estimated [10,27,30]. For this reason, we focused on the $R_0$ range mentioned above, not considering higher values, which would lead to a trivially unstoppable pandemic and thus provide limited information in the analysis of containment scenarios. We considered two dates for the pandemic start that correspond to different seasons in the different hemispheres: beginning of October and beginning of April. Because of the widespread outbreaks among poultry and the number of confirmed human cases of avian influenza in Southeast Asia [1], we considered as potential starting locations both Vietnam and Thailand. We also analyzed scenarios in which the outbreak source is in Europe or in the United States, since pandemics have historically also emerged in these areas [31]. In the Text S1 we report the corresponding results for all the different starting conditions that we considered, including small cities in rural areas where the close proximity between humans and animals might favor the emergence of a novel virus (Text S1). We provide video files representing the geographical spread starting from selected locations (Videos S1–S4).

We analyzed several indicators characterizing the geographical spread, the number of cases, and the time evolution

**Table 1.** Baseline Scenario with No Containment Intervention: Attack Rates and Peak Time in Different Geographical Areas

| Baseline/<br>Reproductive Rate | Region | Average Cases per 1,000 | | Average Arrival Time | Average Peak Time |
| --- | --- | --- | --- | --- | --- |
| | | First year<br>(95% CI) | Second year<br>(95% CI) | Number of Months<br>(95% CI) | Number of Months<br>(95% CI) |
| Baseline | North America | 1.0 (0.1–2.3) | 424 (412–430) | 4 (3–5) | 14.8 (14.5–15.1) |
| $R_0 = 1.5$ | Western Europe | 1.0 (0.5–2.4) | 408 (379–421) | 3 (2–4) | 15.2 (14.9–15.5) |
| | Central America | 145 (5–324) | 390 (388–391) | 5 (3–7) | 13 (11–15) |
| | Southeast Asia | 380 (321–393) | 402 (399–403) | — | 9 (8–10) |
| | Oceania | 409 (378–428) | 412 (380–428) | 4 (3–5) | 8.8 (8.1–9.5) |
| | Lower South America | 159 (9–309) | 185 (71–318) | 6 (4–8) | 10.9 (10.1–11.8) |
| Baseline | North America | 180 (17–357) | 358 (229–503) | 2.5 (1.7–3.3) | 5.8 (5.0–6.6) |
| $R_0 = 1.9$ | Western Europe | 176 (28–324) | 289 (170–505) | 2.4 (1.6–3.2) | 5.8 (5.3–6.3) |
| | Central America | 513 (511–514) | 513 (511–516) | 3.3 (2.5–4.1) | 7.5 (6.6–8.4) |
| | Southeast Asia | 516 (514–518) | 516 (514–522) | — | 5.8 (5.2–6.4) |
| | Oceania | 505 (493–523) | 505 (493–524) | 2.6 (1.8–3.4) | 7.3 (7.2–7.4) |
| | Lower South America | 535 (529–543) | 536 (529–544) | 3.8 (3.0–4.6) | 7.9 (7.5–8.3) |
| Baseline | North America | 549 (475–590) | 549 (475–590) | 2.0 (1.4–2.6) | 4.7 (4.2–5.2) |
| $R_0 = 2.3$ | Western Europe | 538 (455–580) | 538 (455–580) | 1.9 (1.3–2.5) | 4.6 (4.2–5.0) |
| | Central America | 576 (575–577) | 576 (575–577) | 2.7 (2.1–3.3) | 5.8 (5.2–6.4) |
| | Southeast Asia | 581 (579–582) | 581 (580–582) | — | 4.5 (4.0–5.0) |
| | Oceania | 527 (525–532) | 527 (525–532) | 2.0 (1.4–2.6) | 7.0 (6.9–7.1) |
| | Lower South America | 581 (570–590) | 581 (570–590) | 3.0 (2.4–3.6) | 7.1 (7.0–7.2) |

For $R_0 = 1.5$, 1.9, and 2.3 and for six different regions belonging to the northern, southern, and tropical climate zones, we report: (i) the average number of cases per 1,000 at the end of the first and second year; (ii) the average arrival time (i.e., first case detected in the region); and (iii) the average peak time. Numerical simulations start in October with a single symptomatic infected individual in Hanoi.
CI, confidence interval.
doi:10.1371/journal.pmed.0040013.t001

of the pandemics. While we report detailed figures for six main geographical areas as defined in the Text S1 (North America, Western Europe, Central America, Southeast Asia, Lower South America, and Oceania), the simulations provided a detailed analysis at the level of a single country and/or urban area, as reported in the Text S1. The stochastic nature of the model enabled us to measure the probability that an outbreak will evolve to global proportions, thus affecting more than one country, and the corresponding average number of cases that it will generate overall and in each urban area. In Figure 3A we consider a pandemic starting in Hanoi, in the tropical climate zone, and we report for the different reproductive rates the obtained probability that an outbreak will reach a number of countries in the ranges 2–10, 11–50, 51–100, and higher than 100. In the case of a low reproductive rate ($R_0 = 1.1$), approximately 80%–90% of the epidemics starting with a single infectious individual undergo a stochastic extinction due to the low reproductive rate. Moreover, even if the virus spreads outside the initially infected country, the average number of cases is very low (not more than ten per 100,000), so pandemics with reproductive rates close to $R_0 = 1.1$ represent a very mild threat at the global level.

This scenario changed dramatically when we considered higher reproductive rates. For $R_0 \geq 1.5$, the proportion of epidemics reaching a global scale is larger than 40%. For such outbreaks, more than 100 countries are always affected with a global prevalence that reaches values up to 500 symptomatic cases per 1,000. The timeline of the epidemic impact for the regions studied is reported in Figure 3B. For $R_0 = 1.5$ the arrival of the epidemic is strongly dependent on the seasonal effect. In particular, the Northern or Southern hemisphere may or may not experience the epidemic peak during the first year, depending on the season in which the pandemic starts.

In Table 1, along with the pandemic attack rate after the first and second year, we report the average epidemic arrival time (first case detected) and the average epidemic peak time.

The epidemic profile depends on the seasonality and the transportation network, as well as the starting seed. In Figure 4 we report outbreak data (Figure 4A) and the timeline (Figure 4B) for a pandemic starting in Bucharest. We modeled a pandemic originating in Romania, because this was one of the European countries with confirmed presence of the highly pathogenic H5N1 avian influenza virus in domestic birds by the beginning of the year 2006 [1]. In this case the starting seed is in the Northern hemisphere, and results show important differences with the case depicted in Figure 3, regarding both the global impact over one year and the timeline. In particular, the role of seasonality is clear under these conditions. For a starting date in October (i.e., fall in the Northern hemisphere), the probability of global outbreak is very similar to that observed in Figure 3. A difference is found in the occurrence of the epidemic peak in Western Europe within the first year for $R_0 = 1.5$, due to the high traffic connections of this region with Romania. In contrast, if the starting date corresponds to April (i.e., spring in Bucharest), the likelihood of a global outbreak is dramatically reduced, and the corresponding global attack rates are one to two orders of magnitude lower with respect to the scenario in which Hanoi is the initially infected city. The strong decrease in the probability of worldwide outbreaks and in the number of observed cases is due to the fact that the starting conditions—both temporal and geographical—are not in favor of a fast and global spread of the disease, since it originates in the Northern hemisphere during the non-influenza season, thus being characterized by a lower transmission parameter.

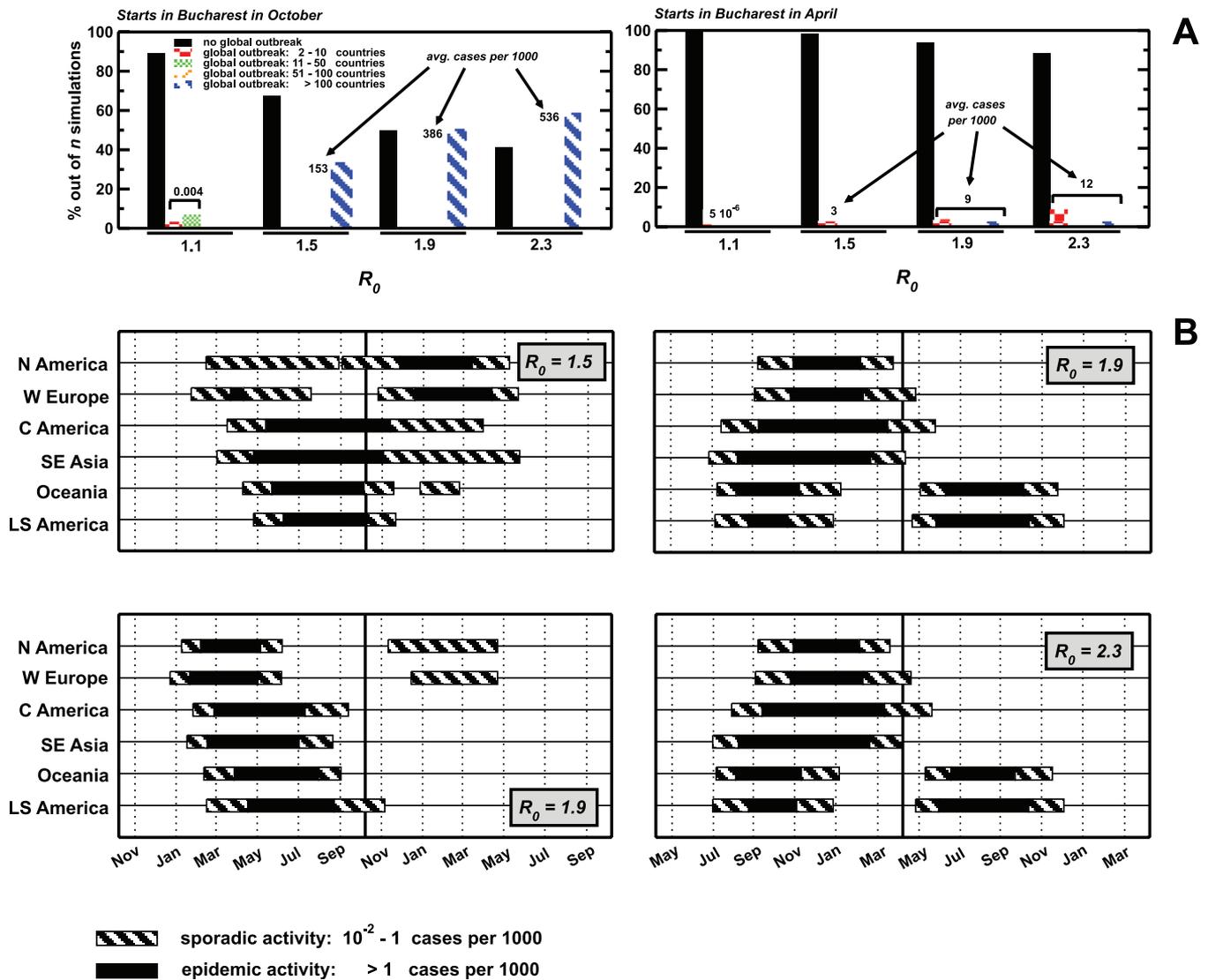

**Figure 4. Baseline Scenario: Probability of Global Outbreaks and Expected Evolution for a Pandemic Seeded in Bucharest**

(A) Probability of observing a global outbreak after one year from the beginning of the epidemic. In the case of outbreaks starting in April, the reproductive rate is effectively reduced to a value smaller than one because of seasonal effects. This effect makes global outbreaks unlikely.
(B) Timing of the influenza pandemic in the six regions under study. Results for different values of the reproductive rates are shown: $R_0 = 1.5$ and $R_0 = 1.9$ for a pandemic starting in October, and $R_0 = 1.9$ and $R_0 = 2.3$ for a pandemic starting in April. Results shown refer to the average profiles in each region when a global outbreak affecting more than 100 countries occurs, thus showing a typical timeline of a pandemic when affecting all regions.
doi:10.1371/journal.pmed.0040013.g004

At intermediate reproductive rates different geographical regions belonging to the same hemisphere may experience a first pandemic wave several months apart, as shown in Figures 3 and 4. This is an effect due to the airline transportation network, which connects different global regions with highly heterogeneous fluxes. At higher reproductive rates, this effect is diminished due to the level of infectiousness of the pandemic, in which seasonal effects and differences in the timescales induced by transportation fluxes are irrelevant.

## Containment Strategies

Scenarios for worldwide pandemic containment or mitigation usually consist of three main factors: vaccination, travel restrictions, and therapeutic and prophylactic use of AV drugs. Additional strategies consider nonpharmaceutical interventions, such as quarantine, early isolation of infectious individuals, school and workplace closures [7,10,11,32,33]. In an emerging pandemic, however, the production and deployment of adequate vaccine supplies would require a time period on the order of six to eight months [2]. Global travel limitations, considered a nonmedical intervention aimed at containing the virus propagation, would be unfeasible since they are so economically disruptive [4,34]. In the Text S1 we show that even drastic travel limitations delay the pandemic evolution by only a few weeks as compared to the baseline case, with almost no impact on the morbidity, as shown elsewhere [32,34] (but see also the controversial aspects discussed in [35–37]). The international community would therefore have to rely initially on AV drugs, which represent a viable response measure to a pandemic in the absence of a vaccine and of other feasible interventions [4,6,8,9,11]. In this context, several recent studies considered the use of AV drugs

**Table 2.** Intervention Scenario with Maximal AV Coverage

| Value of $p_{AV}$ | Value of $R_0$ | Region | Average Cases per 1,000, First Year (95% CI) | Average Arrival Time, Number of Months (95% CI) | Average Peak Time, Number of Months (95% CI) | Average AV Doses per 1,000 | |
|---|---|---|---|---|---|---|---|
| | | | | | | 8 mo (95% CI) | 12 mo (95% CI) |
| $p_{AV} = 0.7$/day | $R_0 = 1.9$ | North America | 0.04 (0.01–0.15) | 4 (2–6) | >12 | 0.007 (4 $10^{-5}$–0.019) | 0.026 (0.008–0.066) |
| | | Western Europe | 0.08 (0.03–0.21) | 4 (2–6) | >12 | 0.01 (2 $10^{-4}$–0.03) | 0.05 (0.02–0.10) |
| | | Central America | 1 (7 $10^{-4}$–9) | 7 (3–11) | >12 | 0.005 (0–0.023) | 0.7 (8 $10^{-4}$–4.0) |
| | | Southeast Asia | 69 (32–105) | — | >12 | 5 (2–8) | 40 (21–58) |
| | | Oceania | 105 (29–159) | 4 (2–6) | 10.8 (10.6–11.0) | 0.21 (0.01–0.61) | 62 (26–90) |
| | | Lower South America | 1 (5 $10^{-3}$–10) | 8 (5–11) | >12 | 0.0006 (0–0.0034) | 0.7 (6 $10^{-3}$–4.8) |
| | $R_0 = 2.3$ | North America | 9 (1–38) | 2.6 (1.7–3.5) | >12 | 5.3 (0.6–17.3) | 5.5 (0.9–17.4) |
| | | Western Europe | 12 (1–43) | 2.5 (1.6–3.4) | >12 | 7.2 (0.9–21.1) | 7.3 (0.9–21.2) |
| | | Central America | 304 (282–313) | 3.6 (2.7–4.5) | 9 (8–10) | 25 (4–60) | 178 (170–183) |
| | | Southeast Asia | 321 (318–322) | — | 7 (6–8) | 134 (111–152) | 188 (187–189) |
| | | Oceania | 367 (358–374) | 2.7 (1.8–3.6) | 8.2 (8.1–8.3) | 73 (60–80) | 216 (211–219) |
| | | Lower South America | 335 (286–355) | 4.1 (3.2–5.0) | 9.5 (8.9–10.1) | 3.8 (0.6–9.7) | 196 (176–207) |
| $p_{AV} = 0.5$/day | $R_0 = 1.9$ | North America | 0.26 (0.05–1.09) | 3 (2–4) | >12 | 0.04 (9 $10^{-4}$–0.16) | 0.13 (0.03–0.39) |
| | | Western Europe | 0.4 (0.1–1.3) | 3 (2–4) | >12 | 0.081 (0.004–0.260) | 0.20 (0.09–0.40) |
| | | Central America | 24.9 (0.2–102.6) | 5 (3–7) | >12 | 0.08 (5 $10^{-6}$–0.33) | 12.3 (0.1–43.1) |
| | | Southeast Asia | 179 (133–208) | — | 10 (9–11) | 16 (7–27) | 89 (73–102) |
| | | Oceania | 248 (204–264) | 4 (2–6) | 9.9 (9.3–10.5) | 2.0 (0.3–5.0) | 124 (109–131) |
| | | Lower South America | 21.3 (0.9–92.1) | 6 (3–9) | 11 (10–12) | 0.007 (0–0.027) | 11 (1–36) |
| | $R_0 = 2.3$ | North America | 38 (4–130) | 2.4 (1.6–3.2) | 5.8 (4.9–6.7) | 19 (3–48) | 19 (3–48) |
| | | Western Europe | 46 (5–126) | 2.3 (1.5–3.1) | 5.7 (5.4–6.0) | 23 (4–52) | 23 (4–52) |
| | | Central America | 367 (364–368) | 3.3 (2.5–4.1) | 8.1 (7.2–9.0) | 74 (29–114) | 183 (182–184) |
| | | Southeast Asia | 371 (369–374) | — | 6.3 (5.6–7.0) | 161 (145–175) | 186 (184–187) |
| | | Oceania | 404 (393–416) | 2.6 (1.7–3.5) | 7.9 (7.8–8.0) | 111 (104–115) | 202 (197–207) |
| | | Lower South America | 406 (398–412) | 3.8 (2.9–4.7) | 8.7 (8.3–9.1) | 19 (5–34) | 203 (200–205) |
| $p_{AV} = 0.3$/day | $R_0 = 1.9$ | North America | 1.6 (0.4–6.3) | 3 (2–4) | >12 | 0.32 (0.01–0.98) | 0.59 (0.25–1.18) |
| | | Western Europe | 1.7 (0.4–10.4) | 3 (2–4) | >12 | 0.52 (0.04–1.89) | 0.7 (0.2–2.0) |
| | | Central America | 218 (59–298) | 4 (3–5) | 11 (10–12) | 1.39 (0.01–5.67) | 81 (33–109) |
| | | Southeast Asia | 316 (300–321) | — | 8 (7–9) | 53 (30–72) | 118 (114–119) |
| | | Oceania | 361 (347–373) | 3 (2–4) | 8.6 (8.2–9.0) | 18 (6–29) | 134 (130–138) |
| | | Lower South America | 217 (63–319) | 5 (3–7) | 10.5 (9.9–11.1) | 0.150 (0.001–0.594) | 81 (33–116) |
| | $R_0 = 2.3$ | North America | 164 (33–314) | 2.3 (1.6–3.0) | 5.6 (5.2–6.0) | 61 (18–108) | 61 (19–108) |
| | | Western Europe | 164 (45–299) | 2.1 (1.4–2.8) | 5.6 (5.0–6.2) | 61 (23–103) | 61 (23–103) |
| | | Central America | 438 (437–440) | 3.0 (2.2–3.8) | 7.1 (6.4–7.8) | 127 (93–150) | 164 (163–165) |
| | | Southeast Asia | 441 (439–443) | — | 5.7 (5.1–6.3) | 162 (157–164) | 165 (164–166) |
| | | Oceania | 446 (435–461) | 2.3 (1.5–3.1) | 7.5 (7.4–7.6) | 124 (117–130) | 166 (162–171) |
| | | Lower South America | 473 (468–481) | 3.5 (2.7–4.3) | 8.0 (7.8–8.2) | 68 (40–88) | 177 (175–180) |

For $R_0 = 1.9$, 2.3, and for different values of the AV distribution rates ($p_{AV} = 0.7$/day, 0.5/day, and 0.3/day) we report for each region: (i) the average number of cases per 1,000 after the first year; (ii) the average arrival time (i.e., first case detected in the region); (iii) the average peak time; and (iv) the average number of AV doses per 1,000 needed after 8 and 12 mo from the start of the pandemic to fully implement the relative treatment program (determined by the rate of treated cases $p_{AV}$). The initial condition is given by a single symptomatic infected individual in Hanoi at the beginning of October. AV treatments start after 20 cases have been identified in at least one country and then with a delay of three days after the identification of the first case in all secondary outbreaks.

CI, confidence interval.

doi:10.1371/journal.pmed.0040013.t002

as an effective strategy for the local containment of the emerging strain at the source [7,10].

In the following, we analyze pandemic containment and mitigation strategies based on the therapeutic use of AV drugs. To provide a comparison of containment scenarios on the global level, we considered the therapeutic treatment of infected individuals with specific AV drugs such as neuraminidase inhibitors [38]. We thus expanded the compartmental model used for the baseline case, as reported in Figure 2. In particular we considered the application of different AV therapeutic protocols as measured by the rate of treatment $p_{AV}$ per symptomatic cases per unit time [6–8]. This parameter models the efficiency and rapidity of case detection and treatment delivery. Finally, we also considered that AV treatment interventions are set up with a characteristic delay [7]. We consider that large-scale AV treatments start after the detection of a cluster of 20 infectious individuals in the country where the epidemic starts and three days after the detection of the first infected case in all countries afterward (a sensitivity analysis to the two different time delays is shown in the Text S1).

## Maximal AV Coverage

At first we investigated the ideal situation in which every country has a stockpile sufficient to treat all cases according to the treatment protocol, thus allowing for the evaluation of the corresponding maximum stockpile level needed. In Table 2, we report the necessary stockpiles (courses per 1,000 people) for each geographical region according to the three AV therapeutic protocols studied (i.e., $p_{AV} = 0.7$/d, 0.5/d, and 0.3/d, corresponding to a probability of 70%, 50%, and 30% per day to have symptomatic cases entering an AV treatment, respectively). We focused on the threatening cases of $R_0 > 1.5$, for a pandemic seeded in Hanoi in October. For $R_0 = 1.9$, AV therapeutic protocols that focus on a timely treatment of

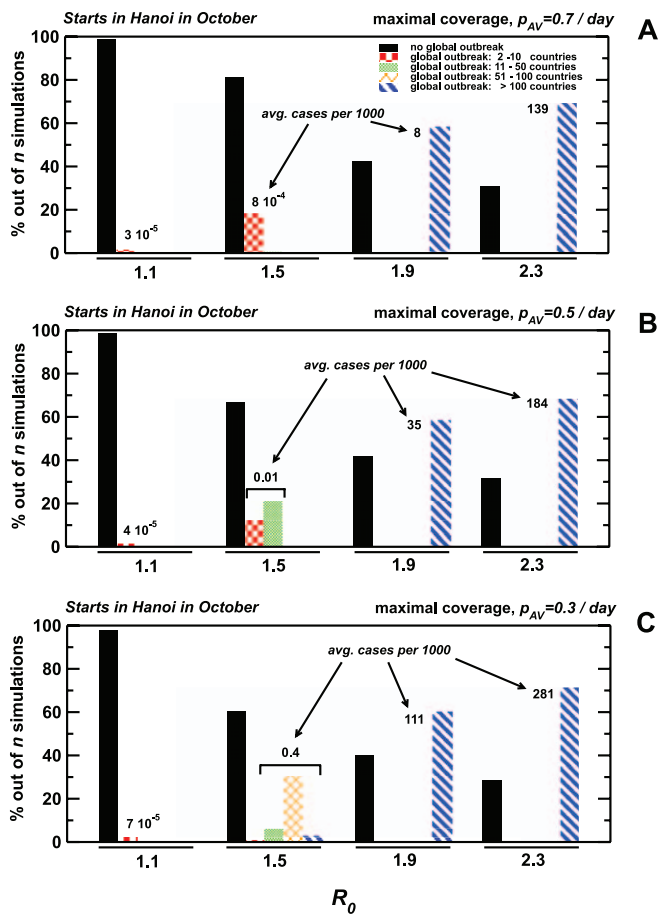

**Figure 5.** Intervention Scenario with Maximal AV Coverage: Probability of Global Outbreaks for a Pandemic Seeded in Hanoi in October

Probability of observing a global outbreak after one year from the beginning of the epidemic, for the values $R_0 = 1.1$, 1.5, 1.9, and 2.3. A maximal coverage with unlimited AV supplies in all hit countries is assumed. Symptomatic cases receive AV drugs with rate $p_{AV} = 0.7/d$ (A), 0.5/d (B), and 0.3/d (C).

doi:10.1371/journal.pmed.0040013.g005

symptomatic cases ($p_{AV} \sim 0.7/d$) appear to be very effective in mitigating the pandemics. These protocols limit the global attack rate to a few cases per 1,000 with the use of AV stockpiles that do not exceed 2%–3% of the world population during the first year, reaching peaks of 4%–6% in certain regions. Moreover, the epidemic peak is almost never observed within the first 12 months. We do not report data for time windows larger than one year, since at that point vaccination intervention should be considered. The possibility of mitigating and delaying a pandemic peak for about one year is strongly supportive of the effectiveness of AV drug use as the first form of intervention. Figure 5 displays the probability of having a global outbreak for different values of $R_0$ for the various therapeutic protocols studied. Its comparison with Figure 3A shows that the application of AV interventions throughout the world strongly affects the probability that the virus will propagate out of the initially infected country. For $R_0 = 1.1$, the probability of containing the outbreak at the source is increased up to 97%–98% in the case of maximal coverage. The probability distribution of the number of countries infected changes dramatically also for $R_0 = 1.5$. The occurrence of a global outbreak may be eliminated because the AV use reduces $R_0$ to close to one in the initially hit countries bursting the probability of random extinction. This is consistent with the results of other studies [7,10] stating that AV interventions are more likely to contain the pandemic at the source. However, as soon as $R_0$ increases, there is a large probability that while the pandemic is mitigated and may appear under control locally, just a few traveling individuals may introduce epidemic seeds in other countries and trigger the global spread of the pandemic (see Figure 6A and 6B referring to two different values of the attack rate at the source). In this context, models that do not consider international travel [33] are missing an important variable and cannot provide a full picture of the occurrence of global pandemics.

In the case of high reproductive rates ($R_0 = 2.3$), the therapeutic use of AV drugs alone can neither contain nor mitigate the pandemics. The number of AV doses that would be needed within one year reaches figures that amount to 20% of the population, with a clinical attack rate up to 30%–50% of the population in the most affected regions. Even in the most optimistic case, such a level of AV intervention would be sustainable only in a few countries for a few initial months, and this costly effort would still not mitigate the pandemic.

## Limited AV Supplies

The results of the previous section suggest that AV therapeutic use can be very effective in a considerable range of virus infectiousness (up to $R_0 = 1.9$) if drugs are available worldwide for a timely use in all countries. Unfortunately, while approximately one-fifth of the world's countries have developed a pandemic response plan [11], fewer than 20 countries are organizing considerable stockpiles of AV drugs [39]. Moreover, it is difficult to foresee an optimal level of supply available worldwide at the start of the pandemic. The recommendation by the World Health Organization is to stockpile drugs in advance, but with the actual rate of production it is reasonable to expect that AV drugs will be neither available in adequate quantities nor distributed homogeneously among countries during the initial stage of the pandemic [11]. For this reason we defined the following different scenarios.

The first scenario assumes that a limited number of countries (Western European countries, the United States, Canada, Australia, New Zealand, and Japan) have stockpiles sufficient to treat 10% of their own population. In addition, we assume that only the two countries initially hit by the pandemic (in our case Vietnam and Thailand) will receive considerable stockpiles (up to 10% of the population) in the attempt to contain the pandemic at the source. With the exception of the two hit countries, AV doses are thus used in an uncooperative way and exclusively on a local basis by those countries that stocked up in preparation for a pandemic.

The second scenario, defined as cooperative, proposes that the prepared countries (listed above) are willing to provide a small fraction of their own supplies, to redistribute them on a worldwide scale where needed. We examine increasingly cooperative strategies in which all prepared countries give up one-tenth (cooperative strategy I) or one-fifth (cooperative strategy II) of their stockpiles for international use. The level of AV coverage in prepared countries thus is lowered to 9%

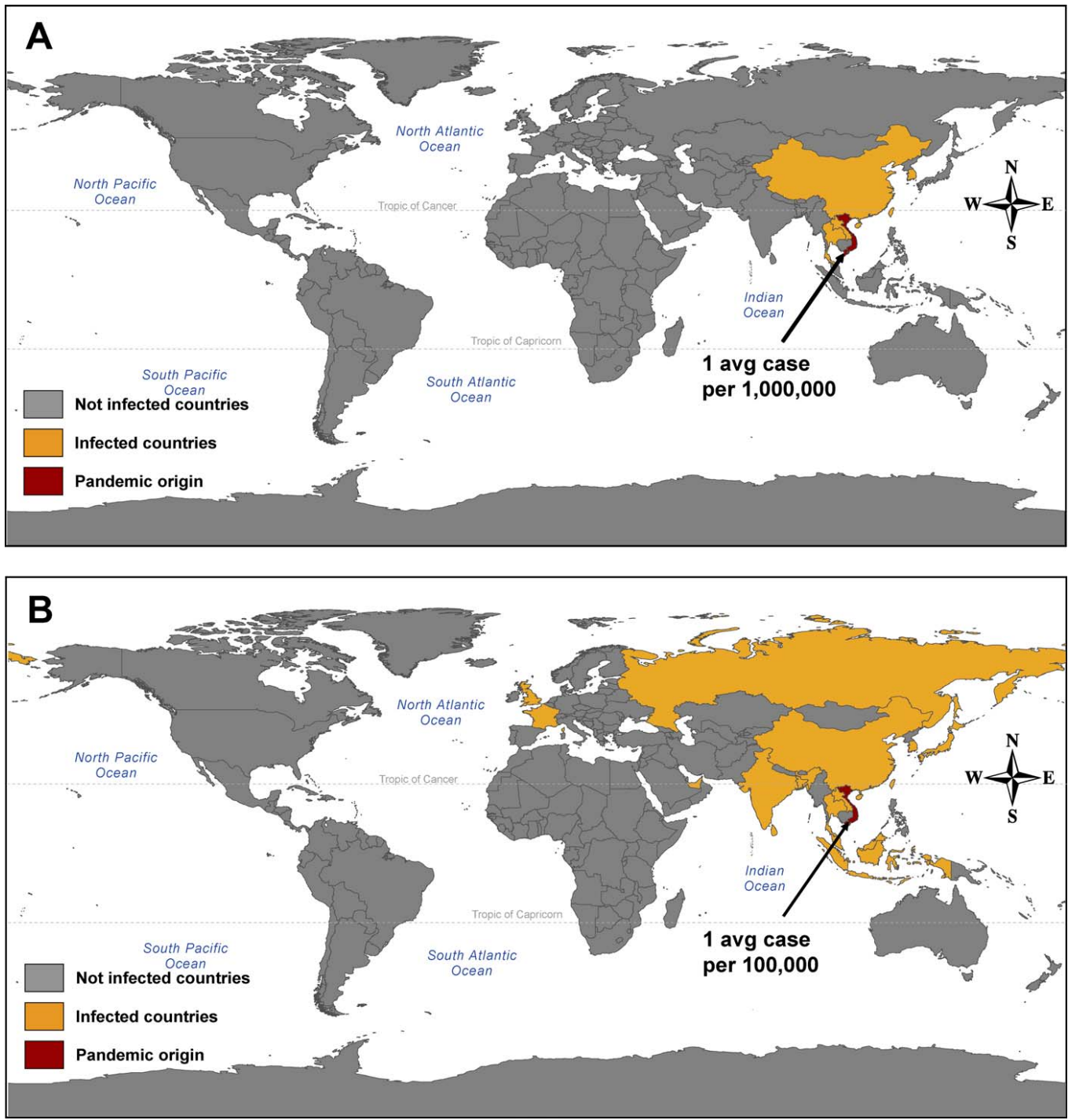

**Figure 6.** Importance of Air Travel in the Worldwide Spread of Pandemic Influenza

The two snapshots show the countries in orange that have a nonnull probability of being infected by the time Vietnam (seeded country) experiences an attack rate of $10^{-6}$ (A) and $10^{-5}$ (B) cases. Results refer to a pandemic originated in Hanoi in October with $R_0 = 1.9$, with the assumption of unlimited AV supplies available. A country is defined as infected (experiencing an outbreak) if at least one generation of secondary cases occurs. Although the number of cases inside Vietnam is very low, the virus has already propagated out of the initial borders to other countries, thus providing several different seeds for the worldwide spread of the disease. Maps are obtained from open source geographic data and plotted with ArcGIS software.
doi:10.1371/journal.pmed.0040013.g006

and 8% in cooperative strategies I and II, respectively. The number of courses provided by the prepared countries is assumed to be collected in a global stockpile to be redistributed worldwide to the countries that experience an outbreak. In the Text S1 we also report simulation results in which the donated courses are instead preemptively deployed in each country proportionally to the population size. The first method is most effective in that it allows the flexible deployment of AV drugs where there is an actual need. Symptomatic cases receive the treatment with rate $p_{AV}$ as long as AV drugs are still available, either from the global stockpile or from the country stockpile. The difference between

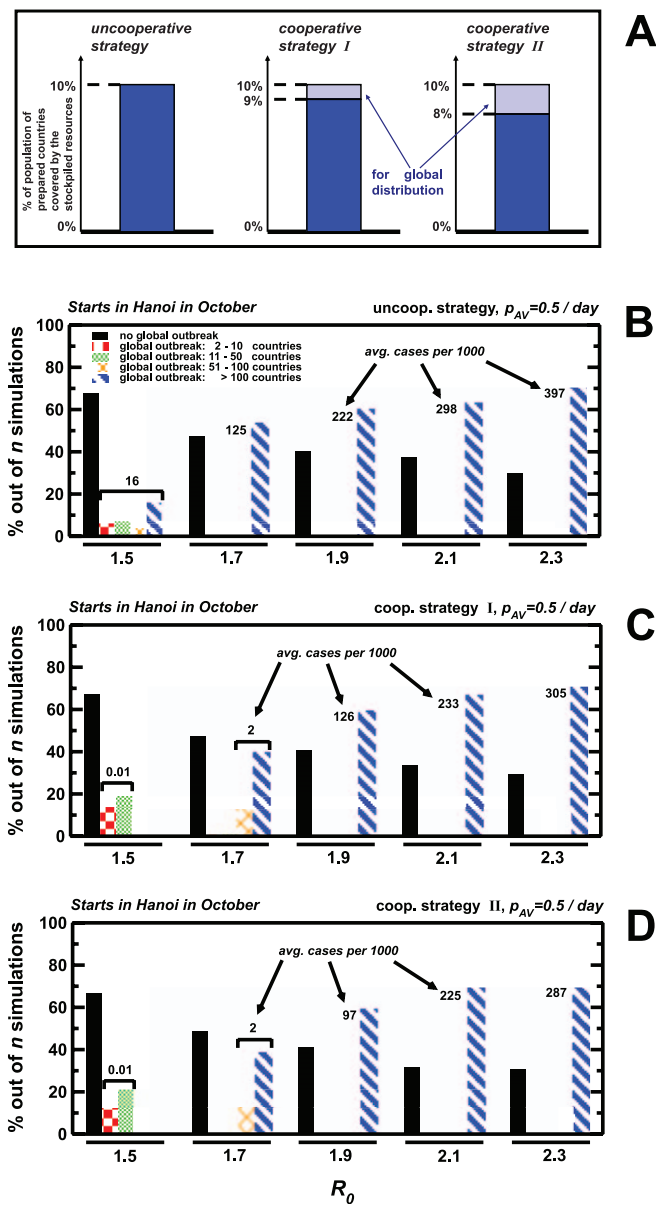

**Figure 7.** Intervention Scenario with Limited AV Supplies: Schematic Representation of the Implemented Containment Strategies and Corresponding Probability of Global Outbreaks, for a Pandemic Seeded in Hanoi in October

(A) The stockpile available in the prepared countries is shown according to the different interventions. An initial stockpile able to cover 10% of the population of prepared countries is assumed. In the uncooperative strategy (left), the totality of these resources is used exclusively for national purposes. The cooperative strategy I (center) is based on the global redistribution of one-tenth of the resources stockpile by prepared countries, which will thus be able to count on supplies covering 9% of their own population. In the cooperative strategy II (right) the amount provided for global sharing is increased to one-fifth, with a corresponding decrease of the supplies from 10% to 8% in the prepared countries stockpiles.

(B–D) Probability of observing a global outbreak after one year from the start of the pandemic in Hanoi, when AV intervention is applied within cooperative or uncooperative strategies. The assumed protocol considers a rate distribution $p_{AV} = 0.5/d$, whenever AV stockpiles are available. The probability of a global outbreak occurring is subdivided into four bins, according to the number of infected countries (see Figure 3A).
doi:10.1371/journal.pmed.0040013.g007

strategies I and II lies in the number of courses available for the unprepared countries. After exhaustion of the global stockpiles, ill individuals in the unprepared countries are no longer treated. Both strategies, however, assume an extensive deployment of AV treatment in unprepared countries from the World Health Organization international stockpile [40] and/or from contributions of other countries, emphasizing the importance of a timely and coordinated response to a pandemic emergency. A representation of the management of resources in the different strategies is shown in Figure 7A.

By focusing on the "threatening cases" with reproductive rates in the range of 1.5–2.3, our results are strikingly in support of a global and cooperative use of AV treatments. To discriminate the effect of the different strategies we must quantitatively analyze the number of symptomatic cases. Figures 7B–7D show that for $p_{AV} = 0.5/d$, in a one-year window, both cooperative strategies outperform the uncooperative one for all reproductive rates. The reproductive rates at which the pandemic is effectively contained worldwide with the cooperative use of AV drugs is obviously changing with the $p_{AV}$ and ranges from $R_0 = 1.9$ for $p_{AV} = 0.7/d$ (Figure 8A and 8B), to $R_0 = 1.7$ for $p_{AV} = 0.5/d$ (Figure 8C and 8D), to $R_0 = 1.5$ for $p_{AV} = 0.3/d$ (Figure 8E and 8F). For values higher than 1.9, the difference between the strategies decreases, but even for a high value such as $R_0 = 2.3$, there is still a difference on the order of 25% in the number of cases between the cooperative and uncooperative strategies, with rapid and efficient drug administration.

This picture is confirmed in Figure 9A in which the profile of symptomatic cases obtained in different global regions for $R_0 = 1.7$ is reported for the baseline case and the modeled intervention scenarios with $p_{AV} = 0.5/d$. The cooperative strategies give rise to one to three orders of magnitude fewer symptomatic cases. In Figure 9B we report the prevalence profile in some countries chosen as illustrative examples, among the six regions considered. Surprisingly, we find that the global coverage intervention is also beneficial in the prepared countries, which benefit from massive stockpiles under any circumstance: the attack rate after the first year is always lower when cooperative strategies are used. This is due to the fact that resources accumulated only in a few countries would not prevent an outbreak explosion in the other regions of the world that are not provided with AV supplies. The inability of mitigating the disease impact on a global level would affect in turn the prepared countries. In addition, global AV interventions give rise to pandemic peaks delayed by more than one year in all world regions. This is extremely important since it indicates that AV treatments distributed on a global scale, even if unevenly, would enable the international community to gain enough time to develop and deploy the appropriate vaccine. Analogous results are obtained also for different initial conditions (see the Text S1).

## Discussion

We have developed a metapopulation computational model that analyzes worldwide scenarios of pandemic outbreaks with different levels of infectiousness and in the presence of different intervention strategies. The study of baseline cases indicates that for low reproductive rates ($R_0 = 1.1–1.3$) the pandemic evolution does not represent a major threat since the attack rate is very limited, and the epidemic

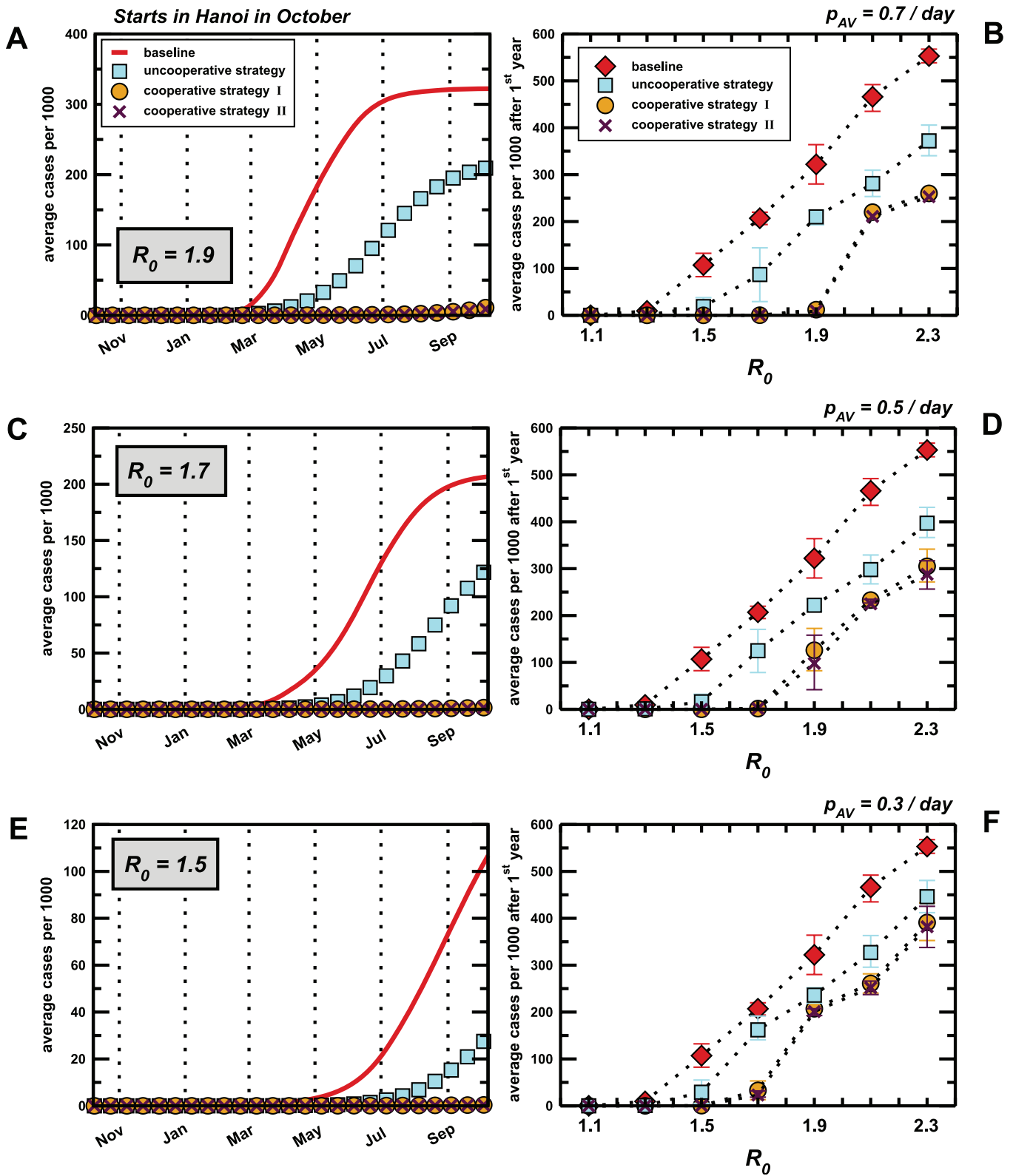

**Figure 8.** Intervention Scenarios with Limited AV Supplies: Global Attack Rates

(A), (C), and (E): Behavior in time of the cumulative number of symptomatic cases per 1,000 people, for a pandemic starting in Hanoi in October, according to different AV repartition strategies and to the baseline case. Given the rate of distribution $p_{AV}$, the value of the reproductive rate $R_0$ up to which cooperative strategies effectively contain the pandemic are reported, $R_0 = 1.9$ and $p_{AV} = 0.7/d$ (A), $R_0 = 1.7$ and $p_{AV} = 0.5/d$ (C), $R_0 = 1.5$ and $p_{AV} = 0.3/d$ (E). (B), (D), and (F): Average number of cases after the first year versus different values of the reproductive rate $R_0$, for a pandemic starting in Hanoi in October. Error bars represent the standard deviation around the average value. Different intervention strategies are compared to the baseline case. Different values of AV distribution rates are shown: $p_{AV} = 0.7/d$ (B), 0.5/d (D), and 0.3/day (F). While cooperative strategies always outperform the uncooperative one, the benefit provided by the redistribution of AV resources decreases as $p_{AV}$ decreases, for fixed values of $R_0$.

doi:10.1371/journal.pmed.0040013.g008

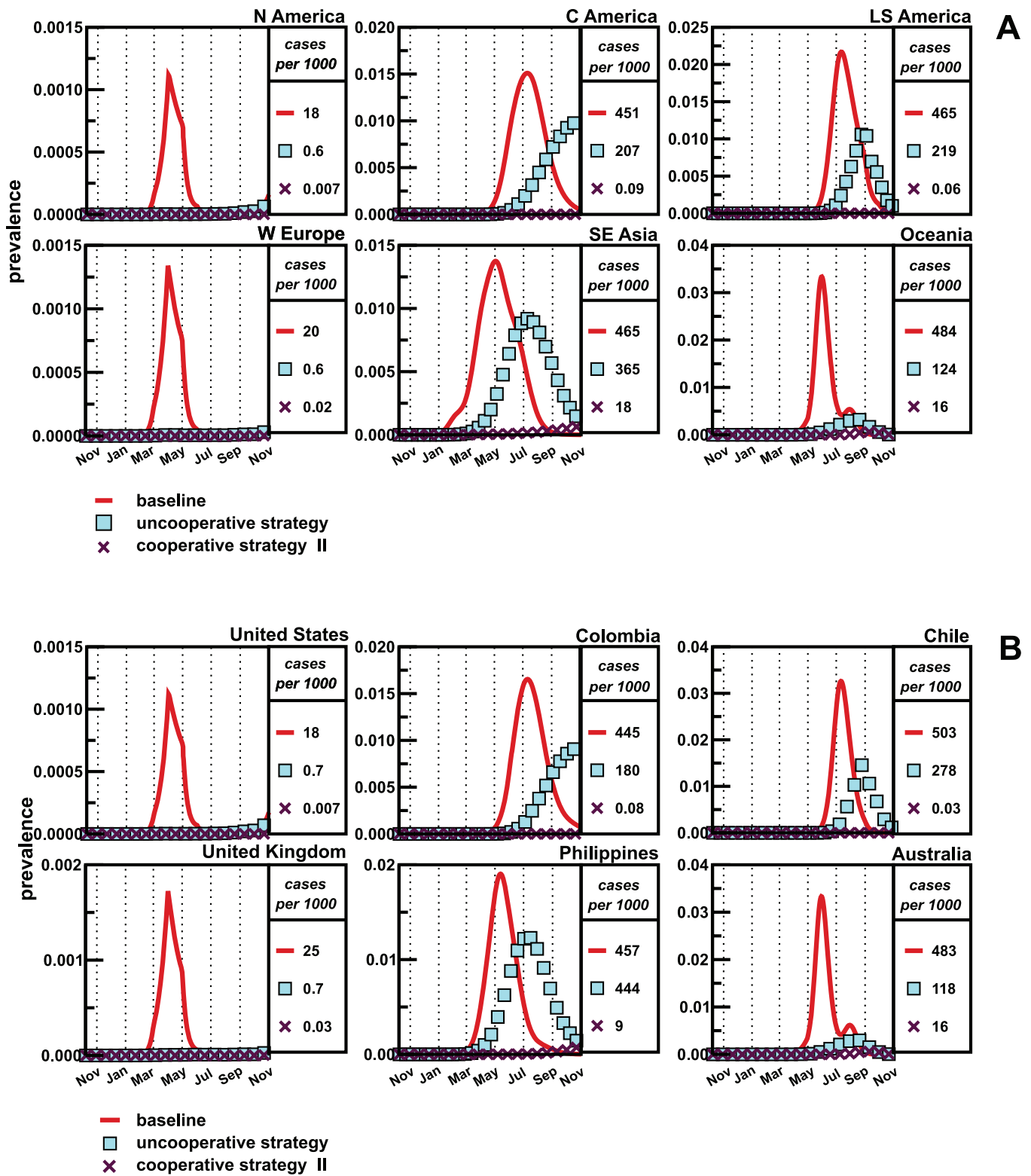

**Figure 9.** Intervention Scenarios with Limited AV Supplies: Expected Pandemic Evolution

Average prevalence profiles expected in the baseline case and the different intervention scenarios under study for a pandemic starting in Hanoi in October. Profiles for six global regions (A) are shown together with six illustrative examples of countries profiles (B), each taken from the corresponding region. For the sake of simplicity, only one cooperative strategy is shown, namely cooperative strategy II. Here $R_0 = 1.7$ and AV drugs, when available, are distributed to symptomatic infectious individuals who enter the AV treatment with a rate $p_{AV} = 0.5$/d. The first 12-month period after the start of the pandemic is shown. The average attack rate after one year is reported for all regions/countries and containment strategies.

doi:10.1371/journal.pmed.0040013.g009

peak is attained well after one year, a time period that allows for the development and deployment of a vaccine. We showed that travel restrictions, which are both economically disruptive and difficult to implement, achieve very modest results, slowing down by only a few days or weeks the overall evolution of the pandemic. Strategies based on AV therapeutic protocols appear, instead, to be a very effective option.

We found that AV therapeutic interventions might be sufficient to mitigate the pandemic for reproductive rates as high as 1.9 if a set of general criteria are met: (i) the therapeutic protocols effectively deliver the treatment to a high proportion of the symptomatic cases (indicatively a rate in the order of 50%–70% per day); (ii) sufficient stockpiles of AV drugs are available, generally corresponding to 2%–6% of the global population; (iii) international stockpiles are managed cooperatively, with partial and timely redistribution of the stockpiles of prepared countries. While mitigation of the pandemic might be achieved by following the previous criteria, the containment of the pandemic is unlikely for $R_0 > 1.5$. Indeed, above those values of the reproductive rate, even if the attack rate is extremely small, infected individuals are traveling with appreciable probability, and pandemics affecting hundred of countries are likely, despite the effective reduction of $R_0$. Finally, for large $R_0$ (indicatively $> 1.9$), interventions based solely on AV therapeutic use are sufficient to neither contain nor mitigate the pandemic, which even in the case of massive AV supplies (approximately 20% of the population) reaches clinical attack rates up to 30%–50% of the population in the most affected regions.

The analysis of the cooperative management of AV supplies in which prepared countries redistribute worldwide even a very limited share of their AV resources, as low as one-tenth of all the national stockpiles, results in a global deceleration of the pandemics whereby the peak is delayed by more than one year for reproductive rates as high as 1.5–1.9. The reproductive rate up to which the cooperative strategy is effective depends on the effectiveness of AV drug distribution (i.e., the rate at which symptomatic individuals are detected and AV courses administered). It is noteworthy that the cooperative strategies are also beneficial for the prepared countries that share a part of their own resources for global distribution. The success of cooperative strategies, however, implies the adoption of a "global" perspective in planning pandemic containment, one calling for a coordinated effort among the international community and the World Health Organization, as opposed to a unilateral strategy in which individual countries rely on their own stockpiles.

The present computational approach is the largest-scale epidemic metapopulation simulation at the worldwide level, to our knowledge. While the model is computationally demanding and detailed in the transportation and census descriptions, all these inputs are determined by available and official transport and demographic data. However, given the scale of the approach, we have adopted model assumptions that are worth addressing. We do not have levels of variations in the virus transmissibility among infected individuals within each compartment [41]. Population heterogeneity, in terms of traveling frequency, is another feature that is neglected. Travel frequency is related to financial and economic status and should be implemented by considering specific compart-ments. Finally, we are mainly concerned with urban areas surrounding airports. This implies neglecting rural areas that might be relevant in less developed countries. While we collected part of the data concerning the previously mentioned socioeconomic features (for instance, wealth distribution), we avoided uncontrolled guesses and preferred not to overload the model with parameters that are not yet data driven. With regard to the etiology of the disease, we considered a given set of parameters for the incubation period, the infectious period, and the asymptomatic proportion. While these parameters find different estimates in the literature, it is possible to study their effect by sensitivity analysis, and in many cases their variation can be absorbed in the change of $R_0$.

In general, the proposed model can be improved by considering other sets of transportation data and more detailed socioeconomic factors such as income and age-dependent traveling probabilities. It can, moreover, be used to test additional intervention measures such as quarantine and AV prophylactic use. Possibly the combination of different interventions might indeed increase their efficiency in mitigating and containing the pandemic within a wider range of reproductive rates. We believe that the present approach complements analogous studies at the regional and national level and might be useful in the assessment of preparedness plans and modeling of emerging disease outbreaks.

## Supporting Information

**Text S1.** Modeling the Worldwide Spread of Pandemic Influenza: Baseline Case and Containment Interventions
Supporting information containing details on the model, scenario analysis and parameter sensitivity analysis.
Found at doi:10.1371/journal.pmed.0040013.sd001 (1.6 MB PDF).

**Video S1.** Baseline Scenario for a Pandemic Starting in Hanoi in October 2006 for $R_0 = 1.7$
Found at doi:10.1371/journal.pmed.0040013.sv001 (7.9 MB WMV).

**Video S2.** Baseline Scenario for a Pandemic Starting in Hanoi in October 2006 for $R_0 = 2.3$
Found at doi:10.1371/journal.pmed.0040013.sv002 (7.2 MB WMV).

**Video S3.** Intervention Scenario for a Pandemic Starting in Hanoi in October 2006 for $R_0 = 1.7$: Uncooperative Strategy
Found at doi:10.1371/journal.pmed.0040013.sv003 (5.7 MB WMV).

**Video S4.** Intervention Scenario for a Pandemic Starting in Hanoi in October 2006 for $R_0 = 1.7$: Cooperative Strategy II
Found at doi:10.1371/journal.pmed.0040013.sv004 (5.4 MB WMV).

## Acknowledgments

We are indebted to J. Lloyd-Smith for useful comments and suggestions during the preparation of the final version of the manuscript. We are grateful to the International Air Transport Association for making the commercial airline flight database available to us. We thank Alessandro Flammini for interesting discussions and useful comments.

**Author contributions.** VC, AB, MB, AJV, and AV designed the study and analyzed the data. VC, AB, MB, AJV, and AV contributed to writing the paper. VC, AB, MB, and AV collected data or did experiments for the study.

## Editors' Summary

**Background.** Seasonal outbreaks (epidemics) of influenza—a viral infection of the nose, throat, and airways—affect millions of people and kill about 500,000 individuals every year. Regular epidemics occur because flu viruses frequently make small changes in the viral proteins (antigens) recognized by the human immune system. Consequently, a person's immune-system response that combats influenza one year provides incomplete protection the next year. Occasionally, a human influenza virus appears that contains large antigenic changes. People have little immunity to such viruses (which often originate in birds or animals), so they can start a global epidemic (pandemic) that kills millions of people. Experts fear that a human influenza pandemic could be triggered by the avian H5N1 influenza virus, which is present in bird flocks around the world. So far, fewer than 300 people have caught this virus but more than 150 people have died.

**Why Was This Study Done?** Avian H5N1 influenza has not yet triggered a human pandemic, because it rarely passes between people. If it does acquire this ability, it would take 6–8 months to develop a vaccine to provide protection against this new, potentially pandemic virus. Public health officials therefore need other strategies to protect people during the first few months of a pandemic. These could include international travel restrictions and the use of antiviral drugs. However, to get the most benefit from these interventions, public-health officials need to understand how influenza pandemics spread, both over time and geographically. In this study, the researchers have used detailed information on air travel to model the global spread of an emerging influenza pandemic and its containment.

**What Did the Researchers Do and Find?** The researchers incorporated data on worldwide air travel and census data from urban centers near airports into a mathematical model of the spread of an influenza pandemic. They then used this model to investigate how the spread and health effects of a pandemic flu virus depend on the season in which it emerges (influenza virus thrives best in winter), where it emerges, and how infectious it is. Their model predicts, for example, that a flu virus originating in Hanoi, Vietnam, with a reproductive number ($R_0$) of 1.1 (a measure of how many people an infectious individual infects on average) poses a very mild global threat. However, epidemics initiated by a virus with an $R_0$ of more than 1.5 would often infect half the population in more than 100 countries. Next, the researchers used their model to show that strict travel restrictions would have little effect on pandemic evolution. More encouragingly, their model predicts that antiviral drugs would mitigate pandemics of a virus with an $R_0$ up to 1.9 if every country had an antiviral drug stockpile sufficient to treat 5% of its population; if the $R_0$ was 2.3 or higher, the pandemic would not be contained even if 20% of the population could be treated. Finally, the researchers considered a realistic scenario in which only a few countries possess antiviral stockpiles. In these circumstances, compared with a "selfish" strategy in which countries only use their antiviral drugs within their borders, limited worldwide sharing of antiviral drugs would slow down the spread of a flu virus with an $R_0$ of 1.9 by more than a year and would benefit both drug donors and recipients.

**What Do These Findings Mean?** Like all mathematical models, this model for the global spread of an emerging pandemic influenza virus contains many assumptions (for example, about viral behavior) that might affect the accuracy of its predictions. The model also does not consider variations in travel frequency between individuals or viral spread in rural areas. Nevertheless, the model provides the most extensive global simulation of pandemic influenza spread to date. Reassuringly, it suggests that an emerging virus with a low $R_0$ would not pose a major public-health threat, since its attack rate would be limited and would not peak for more than a year, by which time a vaccine could be developed. Most importantly, the model suggests that cooperative sharing of antiviral drugs, which could be organized by the World Health Organization, might be the best way to deal with an emerging influenza pandemic.

**Additional Information.** Please access these Web sites via the online version of this summary at http://dx.doi.org/10.1371/journal.pmed. 0040013.

- The US Centers for Disease Control and Prevention has information about influenza for patients and professionals, including key facts about avian influenza and antiviral drugs
- The US National Institute of Allergy and Infectious Disease features information on seasonal, avian, and pandemic flu
- The US Department of Health and Human Services provides information on pandemic flu and avian flu, including advice to travelers
- World Health Organization has fact sheets on influenza and avian influenza, including advice to travelers and current pandemic flu threat
- The UK Health Protection Agency has information on seasonal, avian, and pandemic influenza
- The UK Department of Health has a feature article on bird flu and pandemic influenza